\documentclass[conference]{IEEEtran}
\usepackage{amsmath,amsfonts}
\usepackage{graphicx,amssymb}
\usepackage{algorithmic,algorithm}
\usepackage{amsthm,dsfont}
\usepackage{subfigure,url}
\usepackage[noadjust]{cite}
\usepackage{subfigure}
\theoremstyle{plain}

\newcounter{longequ}[longequ]
\addtocounter{longequ}{8}
\DeclareMathOperator*{\minimize}{minimize}

\renewcommand{\baselinestretch}{0.98}
\begin{document}

\title{Channel Estimation for Intelligent Reflecting Surface-Assisted  Millimeter Wave MIMO Systems\\
}

\author{\IEEEauthorblockN{Tian Lin\IEEEauthorrefmark{1},
Xianghao Yu\IEEEauthorrefmark{2},  Yu Zhu\IEEEauthorrefmark{1}, and Robert Schober\IEEEauthorrefmark{2},}
\IEEEauthorblockA{\IEEEauthorrefmark{1}Fudan University,  China, \IEEEauthorrefmark{2}Friedrich-Alexander-Universit\"{a}t Erlangen-N\"{u}rnberg, Germany\\
Email: \IEEEauthorrefmark{1}\{lint17, zhuyu\}@fudan.edu.cn,  \IEEEauthorrefmark{2}\{xianghao.yu, robert.schober\}@fau.de}}
\maketitle

\begin{abstract}
Intelligent reflecting surfaces (IRSs) are regarded as promising enablers for future millimeter wave (mmWave) wireless communication,  due to their ability to create favorable line-of-sight (LoS) propagation environments. In this paper, we investigate  channel estimation in  downlink IRS-assisted mmWave multiple-input multiple-output (MIMO) systems.   By leveraging the sparsity of mmWave channels, we formulate the channel estimation problem as a fixed-rank constrained non-convex optimization problem. To tackle the non-convexity, an efficient algorithm is proposed by capitalizing on alternating minimization and manifold optimization (MO),  which yields a locally optimal solution.  
Simulation results show that the proposed MO-based estimation (MO-EST) algorithm significantly outperforms two benchmark schemes and demonstrate the robustness of the MO-EST algorithm with respect to imperfect knowledge of the sparsity level of the channels in practical implementations.
\end{abstract}

\section{Introduction}
Due to its enormous potential for overcoming the spectrum crunch, millimeter wave (mmWave) communications has become a promising technology for future wireless cellular systems \cite{2014Ayach}. However, mmWave communication is  vulnerable to blockages due to the  limited scattering effects at mmWave frequencies.  Furthermore, in conventional mmWave communication systems, the propagation environment is uncontrollable, and therefore, the quality of service (QoS) is significantly degraded when the line-of-sight (LoS) links are blocked.

Recently,   intelligent reflecting surfaces (IRSs) have been incorporated into  wireless communication systems, mainly due to their capability of customizing favorable wireless propagation environments \cite{2020Wu}. Equipped with a large number of low-cost \textit{passive} reflective elements, e.g., dipoles and phase shifters,  IRSs enable the adaptation of   wireless propagation environments  with limited power consumption \cite{2019Yu}.  This property of  IRSs can be exploited in  mmWave systems \cite{2020Xiu}.  Specifically, when the direct LoS links between the transceivers are blocked, the IRSs can reflect the incident signals to provide an effective virtual LoS link for mmWave communications.  With well-designed reflecting IRS elements, the communication performance can be further enhanced via programmable and reconfigurable signal reﬂections  \cite{2019Wang, 2019Yu}.  

Nevertheless,  the introduction of IRSs brings new challenges, among which the acquisition of channel state information (CSI)  may be the most demanding task. In particular, in addition to the conventional direct channel between the base station (BS) and the user equipment (UE),  two IRS-assisted channels need to be estimated, i.e., the BS-IRS channel and  IRS-UE channel.  Furthermore, since  radio frequency (RF) chains are not available at the passive IRSs, it is not possible to estimate the two IRS-assisted channels directly by regarding the IRS as a conventional RF chain-driven transceiver. Therefore, the classical channel estimation techniques are not applicable in the newly-emerged IRS-assisted communications systems \cite{2019Wang}. 

Recently, several works have investigated  channel estimation  in IRS-assisted wireless systems \cite{2020Gilder, 2019Wang, 2019Hu, 2020Wan, 2019Chen, 2019Wang2}. The authors of \cite{2019Wang} characterized the minimum pilot sequence length for channel estimation in IRS-assisted multi-user multiple-input single-output (MISO) systems based on the least square (LS) criterion.  A two-timescale estimation scheme was proposed in \cite{2019Hu}, where the high-dimensional  BS-IRS channel and the low-dimensional IRS-UE channel are estimated in a large timescale and a small timescale, respectively. To further reduce the pilot overhead, by exploiting the sparsity of the channels,  compressive sensing techniques were utilized in \cite{2019Chen, 2019Wang2, 2020Wan} to solve the estimation problem. However, the  algorithms proposed in these existing works are only applicable in wireless systems with single-antenna users.  Multiple-input multiple-output (MIMO) systems were  studied in \cite{2020Gilder}, where a channel  estimation algorithm for IRS-assisted systems was developed based on  parallel factor decomposition (PARAFAC). While this  
 approach, designed for sub-6 GHz bands, is also applicable in mmWave MIMO systems,  a significant performance loss is expected as the unique channel characteristics of mmWave MIMO systems are not considered, e.g., the sparsity of  mmWave channels.

In this paper, we propose a novel channel estimation algorithm for IRS-assisted point-to-point mmWave MIMO systems. By exploiting the sparsity of  mmWave channels, we formulate the channel estimation problem as a non-convex optimization problem with fixed-rank constraints. Then, we apply the alternating minimization principle to divide the original problem into two subproblems,  which target the estimation of the BS-IRS channel and the IRS-UE channel, respectively. Finally,  manifold optimization (MO)  is employed to address the non-convex rank constraint and  the subproblems are solved iteratively. The developed algorithm guarantees  monotonic convergence to a locally optimal solution.  Simulations results clearly illustrate the performance improvement of the proposed MO-based estimation (MO-EST) algorithm over two benchmark schemes including the state-of-the-art PARAFAC approach in \cite{2020Gilder}.  We also demonstrate the robustness of the proposed MO-EST algorithm with respect to different channel sparsity levels. 

%  Thanks to the utilization of the sparsity of the channel, a significant gain of more than $5$ dB  can be achieved by our algorithm,  compared to the state-of-the-art approach based on the parallel factor decomposition \cite{2020Gilder}. Simulations also show that the performance gap between the MO-EST algorithm and the compared benchmarks,  markedly becoming larger with the increasing of the sparsity level. 

\emph{Notations:} In this paper, the imaginary unit of a complex
number is denoted by $\jmath=\sqrt{-1}$.
The set of nonnegative integers is denoted by $\mathbb{N}=\{0,1,\cdots\}$. $\mathbb{C}^{m\times n}$ denotes the set of all $m \times n$ complex-valued matrices. Matrices and vectors are denoted by boldface capital and lower-case letters, respectively. 
% Scalars are represented by lower-case letters. 
The $i$-th element of vector $\mathbf{x}$ is denoted by ${x}_i$. $\mathbf{I}_N$ denotes the $N \times N$ identity matrix. $(\cdot)^*$, $(\cdot)^T$, $(\cdot)^H$, $\mathrm{rank}(\cdot)$, $\mathrm{tr(\cdot)}$, $\mathrm{vec(\cdot)}$, and $\|\cdot\|_F$ denote the conjugate, transpose, conjugate transpose, rank, trace, vectorization, and Frobenius norm of a matrix, respectively. 
The Khatri-Rao matrix product is represented by  $\odot$. 
$\Re(\cdot)$ and $\mathbb{E}(\cdot)$ denote the real part of a complex number and expectation, respectively.  $\mathrm{diag}(\mathbf{x})$  is a diagonal matrix with the entries of $\mathbf{x}$ on its main diagonal. $\mathcal{CN}(\mathbf{0}, \mathbf{\Sigma})$ denotes  the  circularly symmetric complex Gaussian distribution with zero mean and  covariance matrix $\mathbf{\Sigma}$.

\section{System Model}

\begin{figure}[!t]
 		\centering
 		\includegraphics[height=1.8in]{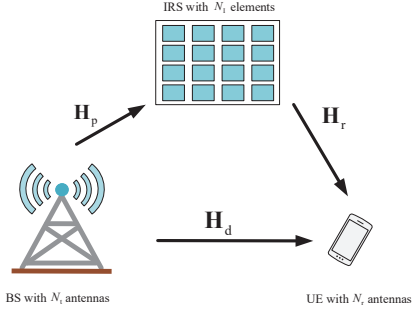}
 	\caption{A downlink IRS-assisted  mmWave MIMO communication system.}
 	    \label{NMSE1}
 	\end{figure}

As shown in Fig. 1, we consider the downlink of an IRS-assisted point-to-point mmWave MIMO communication system.  BS and UE are equipped with uniform  planar arrays (UPAs) consisting of $N_\mathrm{t}$ and $N_\mathrm{r}$ antennas, respectively. In addition, a passive IRS  that employs $N_\mathrm{I}$ phase shifters is deployed in the network to facilitate mmWave communications. 

%  Since the BS and UE are equipped with RF chains, $\mathbf{H}_\mathrm{d}$ can be estimated via conventional algorithms \cite{2019Wang} and thus is assumed known in this paper. 
% 
\subsection{Channel Estimation Protocol}

The channel estimation protocol adopted in this paper is shown in Fig. 2. Specifically, the time available for estimation is divided into $B$ blocks, and each block   consists of $T$ pilot symbol durations. The  reflection coefficient vectors of the IRS  may be  different in different blocks but are constant within one block \cite{2019Chen, 2020Gilder}.    The  pilots received at the UE  in $T$ consecutive time slots  of the $b$-th block, denoted by $\mathbf{R}_b\in \mathbb{C}^{N_\mathrm{r}\times T}$, are compactly written as  
\begin{equation}
    \mathbf{R}_b = \left(\mathbf{H}_\mathrm{r}\mathrm{diag}\left(\mathbf{v}_b\right)\mathbf{H}_\mathrm{p} + \mathbf{H}_\mathrm{d}\right)\mathbf{X}_b + \mathbf{Z}_b,
\end{equation}
where  the BS-UE, BS-IRS, and IRS-UE channel matrices are denoted by  $\mathbf{H}_\mathrm{d} \in \mathbb{C}^{N_\mathrm{r}\times N_\mathrm{t}}$, $\mathbf{H}_\mathrm{p} \in \mathbb{C}^{N_\mathrm{I}\times N_\mathrm{t}}$, and $\mathbf{H}_\mathrm{r} \in \mathbb{C}^{N_\mathrm{r}\times N_\mathrm{I}}$, respectively.  $\mathbf{Z}_b=\left[\mathbf{z}_1,\cdots,\mathbf{z}_T\right] \in \mathbb{C}^{N_\mathrm{r}\times T}$ denotes the received Gaussian noise with $\mathbf{z}_t \sim \mathcal{CN}(\mathbf{0}, \sigma^2\mathbf{I}_{N_\mathrm{r}})$, $\forall t\in\{1,\cdots,T\}$. $\mathbf{v}_b=[v_{b,1},\cdots, v_{b,N_\mathrm{I}}]^T\in\mathbb{C}^{N_\mathrm{I}}$ is the training reflection coefficient vector in the  $b$-th block. Since the IRS is implemented by phase shifters \cite{2019Yu},  the reflecting elements can only change the phases of the received signals, i.e., $|v_{b,n}|=1$. Furthermore, we assume that the  pilot sequences $\mathbf{X}_b \in \mathbb{C}^{N_\mathrm{t}\times T}$ transmitted by the BS are orthogonal to each other, namely, $\mathbf{X}_b\mathbf{X}_b^H=T\mathbf{I}_{N_\mathrm{t}}$ \cite{2020Gilder}. Thus, after removing the pilot symbols at the UE, we have 
\begin{equation}
  \hat{\mathbf{Y}}_b\triangleq \mathbf{R}_b\mathbf{X}_b^H=\left(\mathbf{H}_\mathrm{r}\mathrm{diag}\left(\mathbf{v}_b\right)\mathbf{H}_\mathrm{p} + \mathbf{H}_\mathrm{d}\right) + \mathbf{\hat{Z}}_b,  
\end{equation}
where $\mathbf{\hat{Z}}_b = \mathbf{Z}_b\mathbf{X}^H \in \mathbb{C}^{N_\mathrm{r} \times N_\mathrm{t}}$.

\emph{Remark 1:} By switching off all IRS elements, the direct BS-UE channel can be estimated via traditional algorithms, e.g.,  \cite{2019Wang}. Therefore, in this paper, we assume that $\mathbf{H}_\mathrm{d}$ is known and focus on the estimation of $\mathbf{H}_\mathrm{p}$ and $\mathbf{H}_\mathrm{r}$,  which is the main challenge in channel estimation for IRS-assisted systems. Thus, the relevant part of $\hat{\mathbf{Y}}_b$ is given by
\begin{equation}
    \mathbf{Y}_b =\mathbf{H}_\mathrm{r}\mathrm{diag}\left(\mathbf{v}_b\right)\mathbf{H}_\mathrm{p} + \mathbf{\hat{Z}}_b.
\end{equation}

We  further concatenate the signals received in all $B$ sub-frames as $ \widetilde{\mathbf{Y}}_1 = [\mathbf{Y}_{1}^T,\cdots, \mathbf{Y}_{B}^T]^T \in \mathbb{C}^{BN_\mathrm{r}\times N_\mathrm{t}}$, which leads to
\begin{equation}
     \widetilde{\mathbf{Y}}_1=\left(\mathbf{V}\odot \mathbf{H}_\mathrm{r}\right)\mathbf{H}_\mathrm{p} + \mathbf{\widetilde{Z}}_1,
\end{equation}
where $\mathbf{V} = \left[\mathbf{v}_1,\cdots, \mathbf{v}_B\right]^T \in \mathbb{C}^{B\times N_\mathrm{I}}$ and $\mathbf{\widetilde{Z}}_1 = \left[ \mathbf{\hat{Z}}_1^T,\cdots, \mathbf{\hat{Z}}_B^T\right]^T \in \mathbb{C}^{BN_\mathrm{r} \times N_\mathrm{t}}$.  

\subsection{MmWave Channel Model}
Before formulating the estimation problem, we introduce the channel model for mmWave propagation.  The mmWave propagation environment is well characterized by the Saleh-Valenzuela model \cite{2019Wang2}, which is given by
% In our scheme, we use the first $B$ rows of the $N_\mathrm{I} \times N_\mathrm{I}$  discrete Fourier transform (DFT) matrix to make up the $B \times N_\mathrm{I}$ matrix $\mathbf{V}$.
\begin{figure}[!t]
 		\centering
 		\includegraphics[height=1.8in]{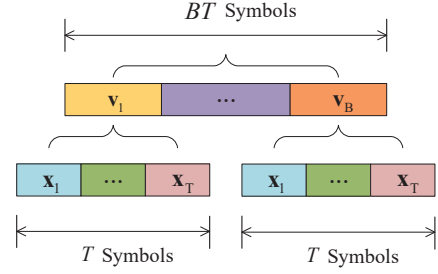}
 	\caption{Frame structure of the channel estimation protocol.}
 	    \label{NMSE1}
 	\end{figure}
\begin{equation}
\label{channel_model}
\begin{split}
 &\mathbf{H}_\mathrm{r} =\sqrt {\frac{{{N_{\rm{r}}}{N_{\rm{I}}}}}{P}} \sum\limits_{p = 1}^P {{\alpha _p}{\mathbf{a}_{\mathrm{r}}}(\theta _\mathrm{r}^p,\phi _\mathrm{r}^p){\mathbf{a}_\mathrm{t}^H}(\theta _\mathrm{t}^p,\phi _\mathrm{t}^p} {)}, \\
 &\mathbf{H}_\mathrm{p} =\sqrt {\frac{{{N_{\rm{t}}}{N_{\rm{I}}}}}{Q}} \sum\limits_{q = 1}^Q {{\beta _q}{\mathbf{a}_{\rm{r}}}(\psi _\mathrm{r}^q,\varphi _\mathrm{r}^q){\mathbf{a}_\mathrm{t}^H}(\psi _\mathrm{t}^q,\varphi _\mathrm{t}^q} {)},
 \end{split}
\end{equation}
where $\alpha_p$, $\theta_\mathrm{r}^p$ ($\phi_\mathrm{r}^p$),  and $\theta_\mathrm{t}^p$ ($\phi_\mathrm{t}^p$) denote the complex gain, azimuth (elevation) angle of arrival (AoA), and azimuth (elevation) angle of departure (AoD) of the $p$-th path of the IRS-UE channel. Similarly, $\beta_q$, $\psi_\mathrm{r}^q$ ($\varphi_\mathrm{r}^q$), and $\psi_\mathrm{t}^q$ ($\varphi_\mathrm{t}^q$) denote  the complex gain, azimuth (elevation) AoA, and azimuth (elevation) AoD of the $q$-th path of the BS-IRS channel. In addition, $\mathbf{a}_\mathrm{r}$ and $\mathbf{a}_\mathrm{t}$ denote the receive and transmit array response vectors, respectively. The  array response vector of a half-wavelength spaced UPA with $M \times N$ elements is given as follows
\begin{equation}
\label{USPA}
\begin{split}
    \mathbf{a}(\theta, \phi) = \frac{1}{\sqrt{MN}}\bigg{[}1,\cdots, e^{\jmath\pi\left(n \sin \theta \sin \phi+m \cos \phi\right)},\cdots, \\
    e^{\jmath\pi\left((\sqrt{N}-1) \sin \theta \sin \phi+(\sqrt{M}-1) \cos \phi\right)}\bigg{]}^T,
    \end{split}
\end{equation}
where $m$ and $n$ are the antenna element indices in the 2-dimensional plane. An important property of mmWave channels is presented in the following lemma.
\newtheorem{lemma}{\textbf{Lemma}}
\begin{lemma} \label{lemma1}
Suppose $\mathrm{min}(N_\mathrm{t}, N_\mathrm{r}, N_\mathrm{I}) \ge \mathrm{max}(P, Q)$, then we have
\begin{equation}
    \mathrm{rank}(\mathbf{H}_\mathrm{r}) = P, \quad \mathrm{rank}(\mathbf{H}_\mathrm{p}) = Q.
\end{equation}
\end{lemma}
\textit{Proof}: Please refer to Appendix A. $\hfill\blacksquare$ 

\emph{Remark 2:} In Section III, the numbers of paths $P$ and $Q$ are  assumed to be  known  at the BS, and hence, the achieved performance is an upper bound for the scenario where $P$ and $Q$ are not available or cannot be accurately estimated. In practice, the numbers of paths can be estimated via  low-complexity compressive sensing methods, e.g., the orthogonal matching pursuit (OMP) method in  \cite{2019Wang2}. In Section IV, we consider the case where $P$ and $Q$ are not exactly known  to evaluate the robustness of the proposed algorithm with respect to a mismatched number of paths. 
\subsection{Problem Formulation}

According to \cite{2019Jen}, the minimum variance unbiased estimators of $\mathbf{H}_\mathrm{p}$ and $\mathbf{H}_\mathrm{r}$  can be obtained based on the LS criterion.  By taking the sparsity of the channels into account and leveraging Lemma 1, we  formulate the channel estimation problem  in IRS-assisted mmWave MIMO systems as follows
\begin{equation}\label{eqn:opt_prob}
\begin{array}{cl}
\displaystyle{\minimize_{\hat{\mathbf{H}}_\mathrm{r}, \hat{\mathbf{H}}_\mathrm{p}}} &  f=\left\| \widetilde{\mathbf{Y}}_1 - \left(\mathbf{V}\odot \hat{\mathbf{H}}_\mathrm{r}\right)\hat{\mathbf{H}}_\mathrm{p}\right\|_F^2 \\
\mathrm{subject \; to} &  \mathrm{rank}(\hat{\mathbf{H}}_\mathrm{r}) = P, \;  \mathrm{rank}(\hat{\mathbf{H}}_\mathrm{p}) = Q,
\end{array}
\end{equation}
where $\hat{\mathbf{H}}_\mathrm{r}$ and $\hat{\mathbf{H}}_\mathrm{p}$ denote the estimates of $\mathbf{H}_\mathrm{r}$ and $\mathbf{H}_\mathrm{p}$, respectively.  Due to the  fixed-rank constraints,   problem (\ref{eqn:opt_prob}) is a highly non-convex problem and a globally optimal solution would entail a very high computational complexity. Besides, the coupling of the two optimization variables in the objective function further complicates the problem.   Thus, in the following, we propose an efficient algorithm that achieves a locally optimal solution of problem (\ref{eqn:opt_prob}).

\section{Proposed MO-EST Algorithm}
To tackle the coupling of the optimization variables in (\ref{eqn:opt_prob}), we first decouple the two variables by applying the alternating minimization principle \cite{2019Lin,2019Yu}. Specifically,  we first fix $\hat{\mathbf{H}}_\mathrm{r}$ and minimize $f$ with respect to the single variable  $\hat{\mathbf{H}}_\mathrm{p}$. The corresponding subproblem is given by
\begin{equation}\label{eqn:opt_prob1}
\begin{array}{cl}
\displaystyle{\minimize_{\hat{\mathbf{H}}_\mathrm{p}}} &  f=\left\| \widetilde{\mathbf{Y}}_1 - \left(\mathbf{V}\odot \hat{\mathbf{H}}_\mathrm{r}\right)\hat{\mathbf{H}}_\mathrm{p}\right\|_F^2 \\
\mathrm{subject \; to} &  \mathrm{rank}(\hat{\mathbf{H}}_\mathrm{p}) = Q.
\end{array}
\end{equation}
To address the non-convex fixed-rank constraint, we apply the MO technique to solve  problem (\ref{eqn:opt_prob1}). Different from traditional compressive sensing methods, e.g., the OMP  and basis pursuit (BP) methods, the proposed MO-based algorithm guarantees  convergence to a locally optimal solution of problem (\ref{eqn:opt_prob1}). 

%\cite{2019Chen, 2019Wang2}

\subsection{Preliminaries of MO}
First, we note that the feasible set of problem \eqref{eqn:opt_prob1} can be represented as a fixed-rank manifold 
\begin{equation}
\mathcal{M}_{Q}\triangleq\left\{\mathbf{X} \in \mathbb{C}^{N_\mathrm{I}\times N_\mathrm{t}}: \operatorname{rank}(\mathbf{X})=Q\right\},
\end{equation}
which is a smooth complex Riemannian manifold. The Riemannian optimization method for the real-valued fixed-rank manifold has been studied in \cite{MO1}. By extending the definitions of the fixed-rank manifold to the complex domain,  we introduce the  key operations that are necessary for the Riemannian optimization method for $\mathcal{M}_Q$.

\textit{1) Inner product:} By endowing the complex space $\mathbb{C}^{N_\mathrm{I}\times N_\mathrm{t}}$ with the Euclidean
metric, the inner product between two points $\mathbf{X}_{1}, \mathbf{X}_{2}\in \mathcal{M}_{Q}$ is defined as
\begin{equation}
    \left\langle \mathbf{X}_{1}, \mathbf{X}_{2}\right\rangle=\Re\left\{\mathrm{tr}(\mathbf{X}_{1}^{H}\mathbf{X}_{2})\right\}.
\end{equation}

\textit{2) Tangent space:} For a point $\mathbf{X}\in \mathcal{M}_{Q}$ on the manifold, its tangent space $T_\mathbf{X} \mathcal{M}_{Q}$, which is
composed of all the vectors that tangentially pass
through $\mathbf{X}$, is given by \cite{MO1}
\begin{equation}\begin{aligned}
T_\mathbf{X} \mathcal{M}_{Q}
\triangleq\{\mathbf{X}_\mathrm{U} \mathbf{M}\mathbf{X}_\mathrm{V}^{H}+ \mathbf{U}_\mathrm{p} \mathbf{X}_\mathrm{V}^{H}+\mathbf{X}_\mathrm{U} \mathbf{V}_\mathrm{p}^{H}: \mathbf{M} \in \mathbb{C}^{Q \times Q}\},
\end{aligned}\end{equation}
where $\mathbf{X}_\mathrm{U}\in\mathbb{C}^{N_\mathrm{I}\times Q}$ and $\mathbf{X}_\mathrm{V}\in\mathbb{C}^{N_\mathrm{t}\times Q}$ denote the semi-unitary matrices containing the first $Q$  left  and right singular vectors of $\mathbf{X}$, respectively. In addition, $\mathbf{U}_\mathrm{p} \in \mathbb{C}^{N_\mathrm{I} \times Q}$ and $\mathbf{V}_\mathrm{p} \in \mathbb{C}^{N_\mathrm{t} \times Q}$ lie in the null spaces of $\mathbf{X}_\mathrm{U}$ and $\mathbf{X}_\mathrm{V}$, respectively, i.e., $\mathbf{U}_\mathrm{p}^{H} \mathbf{X}_\mathrm{U}= \mathbf{0}, \mathbf{V}_\mathrm{p}^{H} \mathbf{X}_\mathrm{V}=\mathbf{0}$.

\textit{3) Orthogonal projection:}  The orthogonal projection of a  point $\mathbf{J}\in \mathbb{C}^{N_\mathrm{I} \times N_\mathrm{t}}$ onto the tangent space of $\mathbf{X}$, $T_\mathbf{X} \mathcal{M}_{Q}$,  is given by
\begin{equation}
    \label{tangent-pro}
    \mathrm{P}_{T_\mathbf{X} \mathcal{M}_{Q}} (\mathbf{J})=  \mathbf{P}_{\mathbf{U}}\mathbf{J} \mathbf{P}_{\mathbf{V}}+\mathbf{P}_{\mathbf{U}}^{\perp} \mathbf{J}  \mathbf{P}_{\mathbf{V}}+\mathbf{P}_{\mathbf{U}} \mathbf{J}   \mathbf{P}_{\mathbf{V}}^{\perp},
\end{equation}
where $\mathbf{P}_\mathbf{U} =\mathbf{X}_\mathrm{U}\mathbf{X}_\mathrm{U}^H$, $\mathbf{P}_\mathbf{V} =\mathbf{X}_\mathrm{V}\mathbf{X}_\mathrm{V}^H$, $\mathbf{P}_{\mathbf{U}}^{\perp}=\mathbf{I}_{N_\mathrm{I}} - \mathbf{P}_\mathbf{U}$, and $\mathbf{P}_{\mathbf{V}}^{\perp}=\mathbf{I}_{N_\mathrm{I}} - \mathbf{P}_\mathbf{V}$. 

\textit{4) Retraction:} Retraction is a mapping from the tangent space to the manifold itself. Particularly, for a point $\widetilde{\mathbf{X}}\in T_\mathbf{X} \mathcal{M}_{Q}$, the retraction operation can be  formulated via a truncated singular value decomposition (SVD)
\begin{equation}
\label{retract}
 \mathcal{R}\left(\widetilde{\mathbf{X}}\right)\triangleq T_{\mathbf{X}}\mathcal{M}_Q\mapsto\mathcal{M}_Q:\widetilde{\mathbf{X}}\mapsto \sum_{i=1}^{Q}\sigma_{i} \mathbf{u}_{i} \mathbf{v}_{i}^{H},
\end{equation}
where $\sigma_{i}$, $\mathbf{u}_{i}$, and $\mathbf{v}_{i}$ are the ordered singular values, left singular vectors, and right singular vectors of $\widetilde{\mathbf{X}}$, respectively.

\subsection{Conjugate Gradient Method on $\mathcal{M}_Q$}
\addtolength{\topmargin}{0.01in}
\begin{figure}[!t]
 		\centering
 		\includegraphics[height=1.8in]{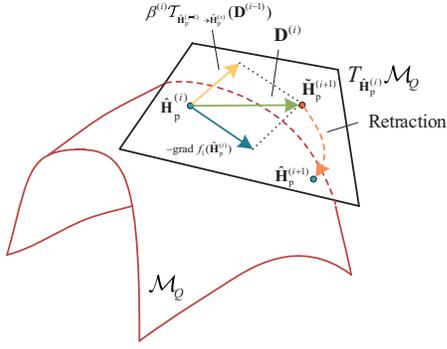}
 	\caption{Illustration of the generalized CG method for the fixed-rank manifold.}
 	    \label{manifold}
\end{figure}
With the basic definitions of the key operations on $\mathcal{M}_Q$ at hand, 
we can formulate the counterpart of the classic conjugate gradient (CG) algorithm in the Euclidean
space  on the manifold $\mathcal{M}_Q$ \cite{MO2, 2019Lin}.
%As  the  objective function of  (\ref{eqn:opt_prob1}) is  quadratic, we devote to generalizing the  well-known conjugate gradient (CG) method  to efficiently perform the optimization on the fixed-rank manifold $\mathcal{M}_{Q}$, which is one kind of Riemannian optimization methods \cite{MO2, 2019Lin}. 
The main idea is illustrated in Fig. \ref{manifold}.  In each iteration, we first find a local minimizer in the tangent space, and then project the obtained point back to the manifold. For problem \eqref{eqn:opt_prob1}, the update rule of the CG method is given by \cite{MO1}
\begin{equation}
\label{search}
    \widetilde{\mathbf{H}}^{(i+1)}_{\mathrm{p}} = \hat{\mathbf{H}}^{(i)}_{\mathrm{p}} + \alpha^{(i)} \mathbf{D}^{(i)},
\end{equation}
where $\alpha^{(i)}$ denotes the Armijo backtracking step size in the $i$-th iteration \cite[Eq. (59)]{MO3}, $\hat{\mathbf{H}}^{(i)}_{\mathrm{p}}$ is the estimate in the $i$-th iteration, and $\widetilde{\mathbf{H}}^{(i+1)}_{\mathrm{p}}$ is the updated local minimizer in the $(i+1)$-th iteration. In addition,  $\mathbf{D}^{(i)}$ is the conjugate direction in the $i$-th iteration, given by
\begin{equation}
\label{direction}
    \mathbf{D}^{(i)} = -\mathrm{grad} f\left(\hat{\mathbf{H}}^{(i)}_{\mathrm{p}}\right) + \beta^{(i)}T_{\hat{\mathbf{H}}^{(i-1)}_{\mathrm{p}} \rightarrow \hat{\mathbf{H}}^{(i)}_{\mathrm{p}}}\left(\mathbf{D}^{(i-1)}\right),
\end{equation}
where the first term is the negative Riemannian gradient representing the   steepest  descent  direction  of  the  objective  function  in  the  tangent  space $T_{\hat{\mathbf{H}}^{(i)}_\mathrm{p}} \mathcal{M}_{Q}$, and $\beta^{(i)}$ represents the chosen Polak-Ribiere parameter \cite[p. 42]{MO3}. 
Since the conjugate direction in the previous iteration $\mathbf{D}^{(i-1)}$ does not lie in $T_{\hat{\mathbf{H}}^{(i)}_\mathrm{p}} \mathcal{M}_{Q}$, the sum operation in \eqref{direction} can not be performed directly. To this end, we introduce the \textit{vector transport} operation to project $\mathbf{D}^{(i-1)}$ to the current tangent space $T_{\hat{\mathbf{H}}^{(i)}_{\mathrm{p}}} \mathcal{M}_{Q}$. According to (\ref{tangent-pro}), the vector transport for $\mathcal{M}_Q$ is given by
\begin{equation}
\begin{split}
 T_{\hat{\mathbf{H}}^{(i-1)}_{\mathrm{p}} \rightarrow \hat{\mathbf{H}}^{(i)}_{\mathrm{p}}}
 = \mathrm{P}_{T_{\hat{\mathbf{H}}^{(i)}_{\mathrm{p}}} \mathcal{M}_{Q}} \left(\mathbf{D}^{(i-1)}\right).
 \end{split}
\end{equation}

Therefore, the remaining task to determine the conjugate direction in \eqref{direction} is to derive the Riemannian gradient.
%can be performed to find a local minimizer in $T_{\hat{\mathbf{H}}^{(i)}_{\mathrm{p}}} \mathcal{M}_{Q}$
%where $\alpha^{(i)}$ denotes the search step size  determined via Armijo backtracking \cite{MO3}.
%
%In order to find a local minimizer of problem (\ref{eqn:opt_prob1}), the Riemannian gradient representing the  steepest increase direction of the objective function in the tangent space needs to be determined. 
Since $\mathcal{M}_Q$ is embedded in $\mathbb{C}^{N_\mathrm{I}\times N_\mathrm{t}}$, the Riemannian gradient is obtained by  projecting the Euclidean gradient  onto the tangent space \cite{MO2}, i.e., 
\begin{equation}
\label{Riemannian-gradient}
    \mathrm{grad} f\left({\hat{\mathbf{H}}_\mathrm{p}}\right)=  \mathrm{P}_{T_{\hat{\mathbf{H}}_\mathrm{p}} \mathcal{M}_{Q}} (\mathbf{G}_1).
\end{equation}
The Euclidean gradient $\mathbf{G}_1$ of $f$ with respect to ${\hat{\mathbf{H}}_\mathrm{p}}$ is given by
\begin{equation}
\label{G1}
  \mathbf{G}_1 = \left( \mathbf{V}\odot \hat{\mathbf{H}}_\mathrm{r}\right)^H\left( \left(\mathbf{V}\odot \hat{\mathbf{H}}_\mathrm{r}\right)\hat{\mathbf{H}}_\mathrm{p} - \widetilde{\mathbf{Y}}_1\right).
\end{equation}
After updating the local minimizer in the $(i+1)$-th iteration according to \eqref{search}, we need to map this minimizer $\widetilde{\mathbf{H}}^{(i+1)}_{\mathrm{p}}$ back to $\mathcal{M}_{Q}$  to obtain the estimate in the $(i+1)$-th iteration, which is achieved by the retraction operation shown in \eqref{retract}, i.e.,
\begin{equation}\label{eq20}
    \hat{\mathbf{H}}^{(i+1)}_{\mathrm{p}}=\mathcal{R}\left(\widetilde{\mathbf{H}}^{(i+1)}_{\mathrm{p}}\right).
\end{equation}
% to complete the CG method for Riemannian manifold and obtain $\hat{\mathbf{H}}^{(i+1)}_{\mathrm{p}}$ via the retraction operation according to  (\ref{retract}).
 The proposed generalized CG method for the fixed-rank manifold,  referred to as the \textbf{CG-MO algorithm},  is summarized in \textbf{Algorithm 1}, where $\epsilon$ is the convergence threshold.

\begin{algorithm}[t]
\label{alg:manifold}
	\caption{CG-MO Algorithm}
	\begin{algorithmic}[1]
			\REQUIRE $\hat{\mathbf{H}}^{(0)}_{\mathrm{p}} \in \mathcal{M}_Q$, $\hat{\mathbf{H}}_\mathrm{r}$, $\mathbf{V}$, $\widetilde{\mathbf{Y}}_1$
	\STATE Set $i=0$ and $f^{(0)}=f\left(\hat{\mathbf{H}}^{(0)}_{\mathrm{p}}\right)$;
\REPEAT	
	\STATE  Compute the Riemannian gradient $\operatorname{grad} f\left(\hat{\mathbf{H}}^{(i)}_{\mathrm{p}}\right)$ according to \eqref{Riemannian-gradient} and \eqref{G1};
	\STATE  Compute the conjugate direction $\mathbf{D}^{(i)}$ according to (\ref{direction});
	\STATE  Update $\widetilde{{\mathbf{H}}}^{(i+1)}_{\mathrm{p}}$ according to (\ref{search}); 
	\STATE  Retract $\widetilde{{\mathbf{H}}}^{(i+1)}_{\mathrm{p}}$ to obtain $\hat{{\mathbf{H}}}^{(i+1)}_{\mathrm{p}}$ according to (\ref{eq20});
    \STATE  $i\leftarrow i+1$;
    \STATE $f^{(i)}=f\left(\hat{\mathbf{H}}^{(i)}_{\mathrm{p}}\right)$;
\UNTIL $f^{(i-1)}-f^{(i)}\le \epsilon$; 
		\STATE Update $\hat{\mathbf{H}}^{(i)}_{\mathrm{p}}$ as the estimate of  ${\mathbf{H}}_{\mathrm{p}}$.
	\end{algorithmic}
\end{algorithm}

\subsection{Estimation of $\mathbf{H}_\mathrm{r}$}
In this subsection, we present the optimization of $\hat{\mathbf{H}}_\mathrm{r}$ for given $\hat{\mathbf{H}}_\mathrm{p}$. First, we establish  the following equality
\begin{equation}
    \left\| \widetilde{\mathbf{Y}}_1 - \left(\mathbf{V}\odot \hat{\mathbf{H}}_\mathrm{r}\right)\hat{\mathbf{H}}_\mathrm{p}\right\|_F^2
    =\left\| \widetilde{\mathbf{Y}}_2 - \left(\mathbf{V}\odot \hat{\mathbf{H}}_\mathrm{p}^T\right)\hat{\mathbf{H}}_\mathrm{r}^T\right\|_F^2,
\end{equation}
where $\widetilde{\mathbf{Y}}_2 = [\mathbf{Y}_{1},\cdots, \mathbf{Y}_{B}]^T \in \mathbb{C}^{BN_\mathrm{t}\times N_\mathrm{r}}$. The subproblem that optimizes $\hat{\mathbf{H}}_\mathrm{r}$ for given $\hat{\mathbf{H}}_\mathrm{p}$ is then formulated as follows
\begin{equation}\label{eqn:opt_prob2}
\begin{array}{cl}
\displaystyle{\minimize_{ \hat{\mathbf{H}}_\mathrm{r}}} &  f = \left\| \widetilde{\mathbf{Y}}_2 - \left(\mathbf{V}\odot \hat{\mathbf{H}}_\mathrm{p}^T\right)\hat{\mathbf{H}}_\mathrm{r}^T\right\|_F^2 \\
\mathrm{subject \; to} &  \mathrm{rank}(\hat{\mathbf{H}}_\mathrm{r}) = P.
\end{array}
\end{equation}
Thus, the \textbf{CG-MO algorithm} is also applicable to solving problem  (\ref{eqn:opt_prob2}).
The main modification compared to the optimization of $\hat{\mathbf{H}}_\mathrm{p}$ is the replacement of the Euclidean gradient in (\ref{G1})  by the Euclidean gradient of $f$ with respect to $\hat{\mathbf{H}}_\mathrm{r}$, which is given by
\begin{equation}
\label{g2}
    \mathbf{G}_2=\left(\hat{\mathbf{H}}_\mathrm{r}\left(\mathbf{V}\odot \hat{\mathbf{H}}_\mathrm{p}^T\right)^T-\widetilde{\mathbf{Y}}_2^T\right)\left(\mathbf{V}\odot \hat{\mathbf{H}}_\mathrm{p}^T\right)^*.
\end{equation}

Finally, the overall estimation scheme  is referred to as the \textbf{MO-EST algorithm} and summarized in \textbf{Algorithm 2}. With the proposed algorithm, the objective values $f$ achieved by the sequence $\left\{\hat{\mathbf{H}}^{(k)}_\mathrm{p}, \hat{\mathbf{H}}^{(k)}_\mathrm{r}\right\}_{k\in \mathbb{N}}$ form a non-increasing sequence that converges to a stationary value, and any limit point of the sequence $\left\{\hat{\mathbf{H}}^{(k)}_\mathrm{p}, \hat{\mathbf{H}}^{(k)}_\mathrm{r}\right\}_{k\in \mathbb{N}}$ is a stationary point of problem (\ref{eqn:opt_prob1}) \cite{MO1}. 

\begin{algorithm}[t]
\caption{MO-EST Algorithm}
\label{MOAlg2}
\begin{algorithmic}[1]
\REQUIRE $\mathbf{V}$, $\widetilde{\mathbf{Y}}_1$, $\widetilde{\mathbf{Y}}_2$
\STATE Randomly initialize $\hat{\mathbf{H}}_{\mathrm{r}}^{(0)} \in \mathcal{M}_P$ and $\hat{\mathbf{H}}_{\mathrm{p}}^{(0)} \in \mathcal{M}_Q$, set $k=0$ and $f^{(0)}=f\left(\hat{\mathbf{H}}_{\mathrm{r}}^{(0)}, \hat{\mathbf{H}}_{\mathrm{p}}^{(0)}\right)$;
\REPEAT	
\STATE $k\leftarrow k+1$;
\STATE Optimize $\hat{\mathbf{H}}_{\mathrm{p}}^{(k)}$ for given $\hat{\mathbf{H}}_{\mathrm{r}}^{(k-1)}$ by solving problem  (\ref{eqn:opt_prob1}) with the CG-MO algorithm;
\STATE Optimize $\hat{\mathbf{H}}_{\mathrm{r}}^{(k)}$ for given $\hat{\mathbf{H}}_{\mathrm{p}}^{(k)}$ by solving problem  (\ref{eqn:opt_prob2}) with the CG-MO algorithm;
 \STATE $f^{(k)}=f\left(\hat{\mathbf{H}}_{\mathrm{r}}^{(k)}, \hat{\mathbf{H}}_{\mathrm{p}}^{(k)}\right)$;
\UNTIL $f^{(k-1)}-f^{(k)}\le \epsilon$; 
\STATE Update $\hat{\mathbf{H}}_{\mathrm{p}}^{(k)}$ and $\hat{\mathbf{H}}_{\mathrm{r}}^{(k)}$ as the estimates of ${\mathbf{H}}_{\mathrm{p}}$ and ${\mathbf{H}}_{\mathrm{r}}$.
\end{algorithmic}\label{alg:manifold}
\end{algorithm}

\section{Simulation Results}

 	 \begin{figure}[!t]
 		\centering
 		\includegraphics[width=2.35in]{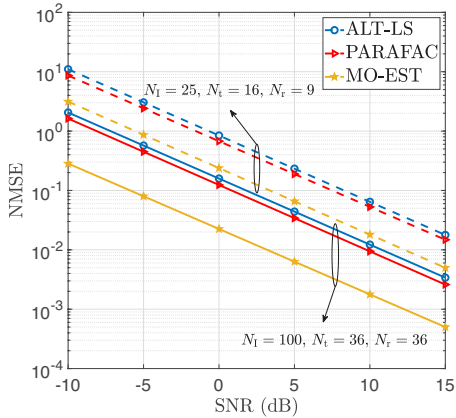}
 	\caption{NMSE versus SNR for different estimation algorithms when $C=3$.}
 	    \label{NMSE1}
 	\end{figure}

In this section, we provide   simulation results for performance evaluation of the proposed MO-EST algorithm.  The signal-to-noise-ratio (SNR) is defined as $\frac{1}{\sigma^2}$. Square UPAs are equipped at both the BS and UE. 
% All results are  averaged over 1000 independent channel realizations. 
For both $\mathbf{H}_\mathrm{p}$ and $\mathbf{H}_\mathrm{r}$, the same number of paths are assumed, i.e., $P=Q\triangleq C$.
According to the channel model in (\ref{channel_model}),  without loss of generality, we let $p = 1$ and $q = 1$ represent the indices of the LoS components in $\mathbf{H}_\mathrm{r}$ and $\mathbf{H}_\mathrm{p}$. The complex channel gains are distributed as $\alpha_1$ ($\beta_1$) $\sim \mathcal{CN}(0,1)$ and $\alpha_i$ ($\beta_i$) $\sim \mathcal{CN}\left(0,10^{-0.5}\right)$ for $i= 2,\cdots, C$ \cite{2019Wang2}. 
The azimuth and elevation AoAs/AoDs, i.e., $\theta_\mathrm{r}^p$ ($\theta_\mathrm{t}^p$) and $\phi_\mathrm{r}^p$ ($\phi_\mathrm{t}^p$), are generated uniformly distributed in $[0$, $\pi]$ and $[-\pi/2$, $\pi/2]$, respectively.  
For the reflecting elements, we set  $N_\mathrm{I}=B$ and use the discrete Fourier transform (DFT) matrix as $\mathbf{V}$. 
The normalized mean square error (NMSE) is adopted as the performance metric. The NMSE is defined as $\mathbb{E}\left\{\|\mathbf{H}_\mathrm{c} - \hat{\mathbf{H}}_\mathrm{c}\|_F^2 / \|\mathbf{H}_\mathrm{c}\|^2\right\}$, where   $\mathbf{H}_\mathrm{c} = \mathbf{H}_\mathrm{r}\mathbf{H}_\mathrm{p}$ and $\hat{\mathbf{H}}_\mathrm{c} = \hat{\mathbf{H}}_\mathrm{r}\hat{\mathbf{H}}_\mathrm{p}$ denote the cascaded channel and its estimate, respectively\footnote{As the $\mathbf{H}_\mathrm{r}$ and $\mathbf{H}_\mathrm{p}$ are coupled in the received signal $\mathbf{Y}_b$, there inevitably exist  scaling ambiguities between $\hat{\mathbf{H}}_\mathrm{r}$ and $\hat{\mathbf{H}}_\mathrm{p}$. Therefore, the NMSE of  $\hat{\mathbf{H}}_\mathrm{c}$ is adopted as  performance metric  to avoid the scaling issues \cite{2020Gilder}.}. The convergence threshold  in both  \textbf{Algorithm 1} and \textbf{2} is set as $\epsilon=10^{-3}$. To show the effectiveness of the proposed MO-EST algorithm, the PARAFAC algorithm  \cite{2020Gilder} is adopted as a benchmark. In addition, by dropping the rank constraints, the LS problems in (\ref{eqn:opt_prob1}) and (\ref{eqn:opt_prob2}) can be alternately solved in closed form. This approach is also adopted as a benchmark and is referred to as the \textbf{ALT-LS algorithm}.  

In Fig. \ref{NMSE1}, the NMSE is plotted as a function of SNR when $C=3$. It can be observed that our proposed MO-EST algorithm  achieves a significant performance gain of more than $4$ dB compared to the two benchmark schemes.   This is mainly because  the proposed MO-EST algorithm exploits the sparsity of the involved mmWave channels. 
In contrast, the ALT-LS algorithm yields the highest NMSE. This phenomenon highlights the importance of incorporating the rank constraints into the alternating optimization algorithm for channel estimation in IRS-assisted mmWave MIMO systems.
Furthermore, as more antennas and reflection elements provide more spatial degrees of freedom for channel estimation, the performance of all three algorithms is improved for larger values  of  $N_\mathrm{t}, N_\mathrm{r}$, and $N_\mathrm{I}$.

 	 \begin{figure}[!t]
 		\centering
 		\includegraphics[width=2.35in]{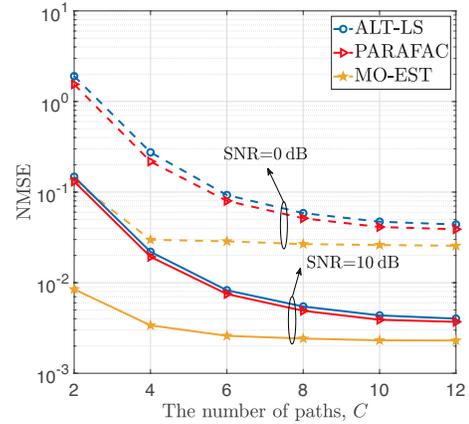}
 	\caption{Effect of the number of paths for different estimation algorithms when $N_\mathrm{r}=16$, $N_\mathrm{t}=36$, and $N_\mathrm{I}=64$.}
 	    \label{NMSE2}
 	\end{figure}
In Fig. \ref{NMSE2}, we investigate the impact of the number of paths of the estimated channels, i.e., $C$,  when $N_\mathrm{r}=16$, $N_\mathrm{t}=36$, and $N_\mathrm{I}=64$. As can be observed,  MO-EST  outperforms the two benchmark algorithms for all considered values of $C$. The performance gain is especially significant in the high sparsity regime. This is  because the performance gain mainly comes from the exploitation of channel sparsity. As the number of paths of the estimated channels increases, the channel sparsity level decreases. Therefore, the performance gap is larger when $C$ is small, which is  typically the case for mmWave channels where scattering is very limited.
% Since the performance improvements are mainly on account of the consideration of the sparsity, the gaps between the MO-EST algorithm and the two benchmark  becomes larger with the smaller number of paths, corresponding to a higher level of sparsity.  
% It should be mentioned that for the scenario that $B$ is very small (e.g. $B<50$), the performance gap between the MO-EST algorithm and the  algorithms becomes larger, which  suggests that the MO-EST algorithm is more suitable for extreme scenarios. 

	 \begin{figure}[!t]
 		\centering
 		\includegraphics[width=2.35in]{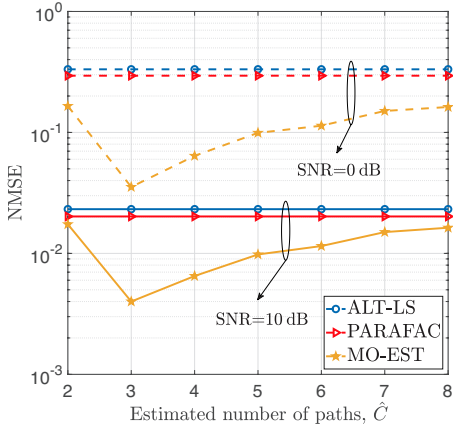}
 	\caption{Effect of  mismatches of $C$ on the NMSE of the MO-EST algorithm when $C=3$, $N_\mathrm{r}=16$, $N_\mathrm{t}=36$, and $N_\mathrm{I}=64$.}
 	    \label{NMSE3}
 	\end{figure}
In Fig. 6, we consider the case where the number of paths, $C$, is not perfectly known for channel estimation and test the robustness of the MO-EST algorithm with respect to the resulting uncertainty. The parameters are set as $C=3$, $N_\mathrm{r}=16$, $N_\mathrm{t}=36$,  $N_\mathrm{I}=64$, and the estimated number of paths is denoted as $\hat{C}$. 
As can be observed, with the proposed MO-EST algorithm, the lowest NMSE is achieved  when $\hat{C}=C$, i.e., the number of paths is perfectly known. In contrast, the performance of the two benchmark algorithms does not depend on  the number of paths, and therefore the achieved NMSEs are independent  of $\hat{C}$.
% We show the NMSE performance of the MO-EST algorithm with different estimations of the number of path, $\hat{P}$ (we assume $\hat{P}=\hat{Q}$) i.e., the estimation $\hat P$ is applied in our proposed MO-EST algorithm while the true value of the number of paths is given by $P$.
% It can be observed that the best NMSE performance is achieved when there is no estimation error in terms of the number of paths, i.e., $\hat{P}=P=3$.
In addition, for the MO-EST algorithm, the mismatch between the estimated $\hat{C}$  and the true value of $C$ leads to a performance loss, which, nevertheless,  is limited especially when $\hat{C}\ge C$. 
In particular, the channel matrix ${\mathbf{H}}_c$ and its estimate $\hat{\mathbf{H}}_c$ can be decomposed by their SVDs, i.e., $\mathbf{H}_c=\Sigma_{c=1}^{C}\sigma_c\mathbf{u}_c\mathbf{v}_c^H$ and $\hat{\mathbf{H}}_c=\Sigma_{c=1}^{\hat{C}}\hat{\sigma}_c\hat{\mathbf{u}_c}\hat{\mathbf{v}}_c^H$, where $\sigma_c$ ($\hat{\sigma}_c$), $\mathbf{u}_c$ ($\hat{\mathbf{u}}_c$), and $\mathbf{v}_c$ ($\hat{\mathbf{v}}_c$) denote the ordered singular values, left singular vectors, and right singular vectors, respectively. In order to minimize  the  objective function in (\ref{eqn:opt_prob}) based on the LS criterion, the MO-EST algorithm  chooses   the ${C}$ largest  singular values of $\hat{\mathbf{H}}_c$ and the corresponding singular vectors to be close to the true values while making the remaining $\hat{C}-C$ singular values small. In other words, the solution obtained by the MO-EST algorithm satisfies $\hat{\sigma}_c \approx {\sigma}_c$, $\hat{\mathbf{u}}_c \approx \mathbf{u}_c$, $\hat{\mathbf{v}}_c \approx \mathbf{v}_c$ for $c=1,\cdots,C$, and $\hat{\sigma}_c \approx 0$ for $c=C+1, \cdots,\hat{C}$, which still maintains a satisfactory estimation performance when $\hat{C}\ge C$. Hence, the proposed MO-EST algorithm is robust with respect to imperfect knowledge of the exact number of paths of the estimated channels. 
% solution minimizing the LS function. Particularly, the  $P$($Q$) largest singular values of the solution is close to  those of  $\mathbf{H}_\mathrm{c}$, while the other $\hat{P}-P$ ($\hat{Q}-Q$)  singular values are tiny, and therefore the MO-EST algorithm exhibits a strong robustness to the mismatches. 

\section{Conclusion}
In this paper, we investigated the channel estimation problem for IRS-assisted  mmWave MIMO systems. By exploiting the sparsity of the mmWave channel, a  manifold optimization-based alternating optimization algorithm, i.e., the MO-EST algorithm, was developed to effectively estimate the BS-IRS  and IRS-UE channels. Simulation results showed the achieved  performance improvements compared to two existing benchmark schemes, even when the sparsity level of the channels was  not accurately unknown.  As a next step, it is of great interest to extend this work to multi-user  and broadband scenarios. 

% \section{Acknowledgements}
% The authors would like to thank Nicolas Boumal at Princeton University
%  for his kind help in the implementation of manifold optimization.

 \begin{appendices}
      \section{}
      According to (\ref{USPA}), the receive array response vector $\mathbf{a}_\mathrm{r}(\theta_\mathrm{r}^p, \phi_\mathrm{r}^p)$  can be written as
\begin{equation}
    \mathbf{a}_\mathrm{r}(\theta_\mathrm{r}^p, \phi_\mathrm{r}^p) = \mathrm{vec}\left( \frac{1}{\sqrt{MN}}\mathbf{\Psi}\left[\bar{\mathbf{a}}^p_{\mathrm{r}, 0},\cdots, \bar{\mathbf{a}}^p_{\mathrm{r}, m},\cdots, \bar{\mathbf{a}}^p_{\mathrm{r}, M-1}\right]^T\right),
\end{equation}
where $\mathbf{\Psi}=\mathrm{diag}\left(\left[ e^{\jmath\pi 0 \cos\phi_\mathrm{r}^p}, \cdots,  e^{\jmath\pi (M-1) \cos\phi_\mathrm{r}^p}\right]\right)$  and $\bar{\mathbf{a}}^p_{\mathrm{r}, m} =  \left[1,\cdots, e^{\jmath\pi (N-1) \sin \theta^p_\mathrm{r} \sin \phi_\mathrm{r}^p}\right]^T$.
% whose adjacent elements are equally proportional. 
When  
 $\mathrm{min}(N_\mathrm{t}, N_\mathrm{r}, N_\mathrm{I}) \ge \mathrm{max}(P, Q)$ , it can be shown that  matrix $\bar{\mathbf{A}}_{\mathrm{r},n} = [{\bar{{\mathbf{a}}}_{\mathrm{r}, m}}^1,\cdots, {\bar{{\mathbf{a}}}_{\mathrm{r}, m}}^P]$ is an $N\times P$ Vandermonde matrix, whose column vectors are linearly independent. Therefore, the vectors ${{\mathbf{a}}}_{\mathrm{r}}(\theta _\mathrm{r}^1,\phi _\mathrm{r}^1),\cdots, {{\mathbf{a}}_{\mathrm{r}}}(\theta _\mathrm{r}^P,\phi _\mathrm{r}^P)$ are also linearly independent and   matrix $\mathbf{A}_{\mathrm{r}}=[{{\mathbf{a}}}_{\mathrm{r}}(\theta _\mathrm{r}^1,\phi _\mathrm{r}^1),\cdots, {{\mathbf{a}}_{\mathrm{r}}}(\theta _\mathrm{r}^P,\phi _\mathrm{r}^P)]$ satisfies 
$\mathrm{rank}(\mathbf{A}_\mathrm{r}) = P$.
 Similarly, $\mathbf{A}_\mathrm{t}=[{{\mathbf{a}}}_{\mathrm{t}}(\theta _\mathrm{t}^1,\phi _\mathrm{t}^1),\cdots, {{\mathbf{a}}_{\rm{t}}}(\theta _\mathrm{t}^P,\phi _\mathrm{t}^P)]$ also  satisfies
$\mathrm{rank}(\mathbf{A}_\mathrm{t}) = P$.
 According to (\ref{channel_model}), $\mathbf{H}_\mathrm{r}$ can be expressed as
 \begin{equation}
 \label{Hr2}
     \mathbf{H}_\mathrm{r} = \mathbf{A}_\mathrm{r}\Sigma \mathbf{A}_\mathrm{t}^H,
 \end{equation}
 where $\Sigma = \mathrm{diag}(\alpha_1,\cdots, \alpha_P)$ is also a rank-$P$ matrix. We have the following inequalities 
 \begin{equation}
 \begin{split}
 \label{rank-pro}
     \mathrm{rank}(\mathbf{A} \mathbf{B}) \ge \mathrm{rank}(\mathbf{A})+\mathrm{rank}(\mathbf{B})-k, \\ \mathrm{rank}(\mathbf{A} \mathbf{B}) \le \min \{\mathrm{rank}(\mathbf{A}), \mathrm{rank}(\mathbf{B})\}
      \end{split}
 \end{equation}
 for arbitrary matrices $\mathbf{A}\in \mathbb{C}^{m\times k}$ and  $\mathbf{B}\in \mathbb{C}^{k\times n}$. Combining the results in (\ref{Hr2}) and (\ref{rank-pro}), it can be shown that 
 \begin{equation}
    \mathrm{rank}(\mathbf{H}_\mathrm{r}) = P,
\end{equation}
and similarly we can prove $\mathrm{rank}(\mathbf{H}_\mathrm{p}) = Q$.
  \end{appendices}

\end{document}